\documentclass[pra,aps,10pt,twocolumn]{revtex4-1}
\usepackage{graphicx}\usepackage[colorlinks=true,linkcolor=blue,citecolor=red,urlcolor=magenta]{hyperref}
\usepackage{amsmath}
\usepackage{color}
\usepackage{times}
\usepackage[top=1.8cm,left=2cm,right=2cm,bottom=2cm]{geometry}

\begin{document}

\title{Comment on ,,Do Bloch waves interfere with one another?''}

\author{Tomasz Sowi\'{n}ski}
\affiliation{\mbox{Institute of Physics, Polish Academy of Sciences, Aleja Lotnik\'{o}w 32/46, PL-02668, Warsaw, Poland}}

\begin{abstract}
We point out that argumentation presented in [Phys. Lett. A {\bf 417}, 127699 (2021)], leading to the conclusion that in periodic systems there is a superselection principle forbidding two different Bloch states to form a coherent superposition of a fixed phase, is unjustified and false. As an example, we show that the operator projecting to the selected Wannier state can be experimentally utilized to determine relative phase between superposed Bloch states. In this way we argue that, in fact, the non-existence of the aforementioned superselection principle is manifested always when localized measurements are performed, for example in state-of-the-art experiments with ultracold atoms confined in optical lattices.\\ 

\end{abstract}
 
\maketitle
\vspace{-1.5cm}
In a recent paper by Vivek M. Vyas \cite{2021VivekPhysLettA} the author claims that energy eigenstates of periodic quantum systems $|\psi_{nk}\rangle$ (where $n$ and $k$ denote band index and wavevector, respectively) characterized by different wavevectors $k$ cannot be coherently superposed since this is forbidden by a specific superselection principle being a straightforward consequence of translational symmetry. Namely, the author argues that for any well-defined hermitian operator $\hat{\cal O}$ all matrix elements $\langle\psi_{nk}|\hat{\cal O}|\psi_{mk'}\rangle$ can be non-vanishing only for $k=k'$ (band indices $n$ and $m$ are arbitrary). If this statement were true then in fact there would be no physical observable giving a chance to determine relative phase $\phi$ in the superposition $|\psi\rangle =\left(|\psi_{nk_1}\rangle + \mathrm{e}^{i\phi}|\psi_{mk_2}\rangle\right)/\sqrt{2}$ and consequently the superposition cannot be distinguished from an incoherent classical mixture described by the density matrix $\rho = \left(|\psi_{nk_1}\rangle\langle\psi_{nk_1}|+|\psi_{mk_2}\rangle\langle\psi_{mk_2}|\right)/2$~\cite{2007BartlettRMP}. However, the author's argument is based on the overly rigorous and unjustified conviction that the translational invariance of the whole system implies that every well-defined hermitian operator acting in the Hilbert space spanned by Bloch functions must be cell-periodic. In fact, cell-periodicity is only a sufficient but for sure not necessary condition to assure a consistent definition of the hermitian operator. Fortunately, abandoning unjustified requirement of the cell-periodicity condition of hermitian operators opens the way to physical manifestation of the relative phase between two Bloch states forming a coherent superposition. 

To give an appropriate and clear example let us consider a one-dimensional lattice system of $N$ cells with periodic boundary conditions and focus on a specific superposition of all Bloch states belonging to a given band $n$:
\begin{equation} \label{Wannier}
|W^{n}_R\rangle = \frac{1}{\sqrt{N}}\sum_k \mathrm{e}^{-ikR}|\psi_{nk}\rangle,
\end{equation} 
with $R$ being an integer multiple of a lattice cell-width $a$ and pointing to a selected lattice site. Here, summation runs over all $k_\ell=2\pi\ell/Na$ with $\ell=0,1,\ldots,N-1$. The superposition \eqref{Wannier} is a well-known Wannier function describing particle with band index $n$ localized in a given lattice site $R$~\cite{1937WannierPhysRev,1962WannierRMP}. The set of all Wannier functions form a complete and orthogonal basis in the Hilbert space, {\it i.e.}, $\langle W^{n}_R|W^{n'}_{R'}\rangle=\delta_{RR'}\delta_{nn'}$ and $\sum_{nR}|W^{n}_R\rangle\langle W^{n}_{R}|=1\!\!1$. Now, let us consider the hermitian operator $\hat{\cal O}=|W^{n_0}_{R_0}\rangle\langle W^{n_0}_{R_0}|$ projecting to a selected Wannier state from band $n_0$ localized on site $R_0$. Of course it can be easily expressed in the basis of Bloch functions
\begin{equation}
\hat{\cal O}=\frac{1}{N}\sum_{kk'} \mathrm{e}^{-iR_0(k-k')}|\psi_{n_0k}\rangle\langle \psi_{n_0k'}|
\end{equation}
Thus, there is no doubt that the operator is well-defined when acting on any Bloch state. Moreover, since all Bloch states form a complete and orthogonal basis in the system's Hilbert space, the hermitian operator $\hat{\cal O}$ is well-defined when acting to any allowed state of the system. At the same time, however, the operator $\hat{\cal O}$ {\sl is not} cell-periodic. Therefore, it can be utilized to determine a relative phase between two Bloch states being superposed. Indeed, it is very easy to check that
\begin{equation}
\langle \psi_{n_0l}|\hat{\cal O}|\psi_{n_0l'}\rangle = \frac{1}{N} \mathrm{e}^{-R_0(l-l')}\neq 0.
\end{equation}  
Thus, the off-diagonal contribution to the interference pattern is essentially non-zero and in principle can be determined by appropriate measurement in the basis of Wannier functions. 

At this point it should be noted that the basis of Wannier functions is a natural basis for genuine measurements performed in experiments with ultracold atoms confined in optical lattices~\cite{2015CheukPRL,2015HallerNaturePhys,2015ParsonsPRL}. Indeed, quantum-gas microscopes allow to determine distinct lattice sites occupied with individual particles, {\it i.e.}, particles being in quantum states described almost perfectly by corresponding Wannier functions. Thus, such physical measurements and their theoretical representations by projectors $\hat{\cal O}$ are natural candidates to determine relative phases between Bloch functions.   

Let us mention that the problem of superposition of Bloch states belonging to Hilbert subspaces characterized by different lattice vectors $k$ has its counterparts in many physical contexts. One of the simplest examples comes from quantum electrodynamics and is related to the problem of superposition of photons of different frequencies. It is well known that the quantum electromagnetic field can be viewed as a system of decoupled harmonic oscillators of different frequencies~\cite{1975BialynickiQED}. Thus, one may wonder whether photons of different energies can in fact coherently interfere. Although this question is experimentally challenging, it has recently been shown in beautiful experiments that such superpositions can not only be engineered but also controlled and exploited for quantum computing~\cite{1999MerollaPRL,2016ClemmenPRL}.

To conclude, presented argumentation shows that relative phase between two (or more) superposed Bloch functions has experimental importance and can be captured by appropriately tailored measurement. Therefore, the answer to the leading question from the original paper~\cite{2021VivekPhysLettA} on whether Bloch waves can interfere at all is as positive as possible while claiming that periodic systems have additional super-selection principle immanently build-in is not justified.

\bibliography{biblio}

\end{document}